\documentclass[vecphys]{svmult}

\usepackage{makeidx}         
\usepackage{graphicx}        
\usepackage{amssymb}                        
\usepackage{multicol}        
\usepackage{cite}            
\usepackage[bottom]{footmisc}

\def\be{\begin{equation}}
\def\ee{\end{equation}}
\def\bit{\begin{itemize}}
\def\eit{\end{itemize}}
\def\bea{\begin{eqnarray}}
\def\eea{\end{eqnarray}}

\def\nd{{\vphantom{\dagger}}}

\def\half{{1\over 2}}

\def\vpi{{\vec \pi}}
\def\v0{{\vec 0}}

\def\bS{{\bf S}}

\def\bR{{\bf R}}

\def\bk{{\bf k}}

\def\ok{{\omega_\bk}}
\def\gk{{\gamma_\bk}}

\def\cA{{\cal A}}

\def\cO{{\cal O}}
\def\cH{{\cal H}}

\def\cV{{\cal V}}

\def\ij{{\langle ij\rangle}}

\def\uar{\uparrow}
\def\dar{\downarrow}
\def\nd{^{\vphantom{\dagger}}}
\def\yd{^\dagger}
\def\frac#1#2{{\textstyle{#1 \over #2}}}
\def\ssr#1{{\scriptscriptstyle{\rm #1}}}
\def\na{n\nd_b}
\def\xk{{\xi\nd_\bk}}
\def\xks{{\xi^*_\bk}}
\def\gks{{\gamma^*_\bk}}

\def\ok{{\omega\nd_\bk}}
\def\gk{{\gamma\nd_\bk}}
\def\bR{{\bf R}}
\def\bxh{{\hat{\bf x}}}
\def\byh{{\hat{\bf y}}}
\def\bzh{{\hat{\bf z}}}
\def\cV{{\cal V}}
\def\Jpa{J\nd_\parallel}
\def\Jpe{J\nd_\perp}
\def\Qpa{Q\nd_\parallel}
\def\Qpe{Q\nd_\perp}
\def\gpa{\gamma\nd_\parallel}
\def\gpe{\gamma\nd_\perp}
\def\rhpa{\rho\nd_\parallel}
\def\rhpe{\rho\nd_\perp}
\def\intl{\int\limits_{-1}^1}

\makeindex             

\begin{document}

\title{Schwinger Bosons Approaches to Quantum Antiferromagnetism}

\author{Assa Auerbach  and  Daniel P. Arovas }
\institute{AA: Physics Department, Technion,\\ Haifa 32000, Israel.\\
\texttt{assa@physics.technion.ac.il}\\
DPA: Department of Physics, University of California at San Diego, \\La Jolla, California 92093, USA\\
\texttt{darovas@ucsd.edu}}


\maketitle
\section{$\textsf{SU}(N)$ Heisenberg Models}
\label{ChapSUN}
The use of large $N$ approximations \index{large $N$ approximations} to treat strongly interacting quantum systems been
very extensive in the last decade. The approach originated in elementary particles theory,
but has found many applications in condensed matter physics.  Initially, the large $N$
expansion was developed for the  
Kondo and Anderson  models of magnetic impurities in metals. Soon
thereafter it was  extended to the Kondo and Anderson lattice models
for mixed valence and heavy fermions phenomena in rare earth compounds\cite{Piers,RN}.  

In these notes we shall formulate and apply the large $N$ approach to the quantum
Heisenberg model\cite{AA,AA2,book,RS1}. This method provides an additional avenue
to the static and dynamical correlations of quantum magnets. The mean field theories
derived below can describe both ordered and disordered phases, at zero
and at finite temperatures, and they complement the semiclassical 
approaches.

Generally speaking, the parameter $N$ labels an internal $\textsf{SU}(N)$  symmetry at each lattice  site
(i.e., the number of ``flavors''  a Schwinger boson or a constrained fermion can have).  In  most
cases, the large $N$ approximation has been applied to treat  spin Hamiltonians, where the
symmetry is \textsf{SU}(2), and $N$ is therefore not a truly large parameter.  Nevertheless, the $1/N$
expansion provides an easy method for obtaining simple mean field theories. These have been
found to be surprisingly successful as well.

The large $N$ approach  handles strong
local interactions in terms of constraints. It is not a perturbative expansion in the size of
the interactions but rather a saddle point expansion which usually preserves the spin
symmetry of the Hamiltonian. The Hamiltonians are written as a sum of  terms $\cO\yd_{ij} \cO\nd_{ij}$, which are biquadratic in the
Schwinger boson creation and annihilation operators,  on each bond on the lattice. This sets up a natural mean field decoupling scheme
using one complex Hubbard Stratonovich field per bond.

At the mean field level, the constraints are enforced only on
average. Their effects are systematically reintroduced by the higher-order corrections in
$1/N$. 

It turns out that
different large $N$ generalizations are suitable for different Heisenberg models, depending on
the sign of couplings, spin size, and lattice. Below, we describe two large $N$ generalizations  of the Heisenberg antiferromagnet.

\section{Schwinger Representation of $\textsf{SU}(N)$ Antiferromagnets}
The \textsf{SU}(2) algebra is defined by the familiar relations $[S^\alpha,S^\beta]=i\epsilon_{\alpha\beta\gamma}S^\gamma$.
The spin operators commute on different sites, and admit a bosonic representation.   Since the spectrum of a bosonic
oscillator includes an infinite tower of states, a constraint is required in order to limit the local Hilbert space dimension
to $2S+1$.  In the Holstein-Primakoff representation, one utilizes a single boson $h$, writing 
$S^+=h^\dagger\sqrt{2S-h^\dagger h}$, $S^-=(S^+)^\dagger$, and $S^z=h^\dagger h-S$, together with the non-holonomic
constraint $0\le h^\dagger h \le 2S$.  The square roots prove inconvenient, and practically one must expand them as a power
series in $ h^\dagger h /2S$.  This generates the so-called spin-wave expansion.

\begin{figure}[t]
\centering
\includegraphics*[width=7cm]{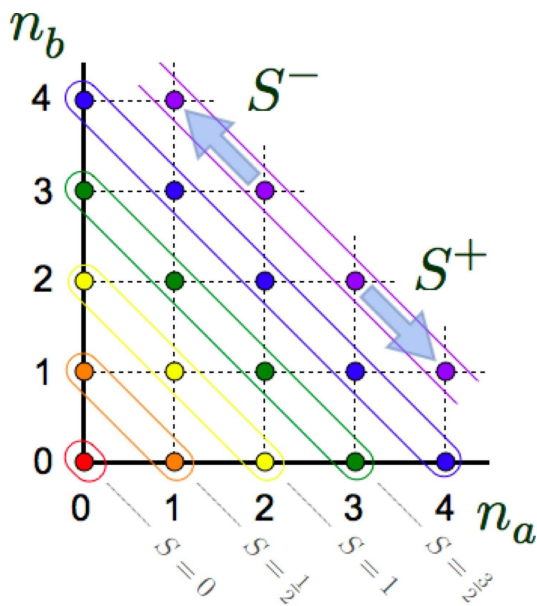}
\caption{\label{sbrep} Schwinger representation of \textsf{SU}(2).} 
\end{figure}

Another representation, due to Schwinger, makes use of two bosons, $a$ and $b$.   We write
\be
S^+=a^\dagger b \quad,\quad S^- = b^\dagger a \quad,\quad S^z=\frac{1}{2}(a^\dagger a - b^\dagger b)\ ,
\label{sbdef}
\ee
along with the holonomic constraint,
\be
a^\dagger a + b^\dagger b=2S\ ,
\label{sbcon}
\ee
where the boson occupation, $2S$, is an integer which determines the representation of \textsf{SU}(2).  This scheme is
depicted graphically in fig. \ref{sbrep}.

There are three significant virtues of the Schwinger representation.  The first is that there are no square roots to expand.
The second is that the holonomic constraint (\ref{sbcon}) can be elegantly treated using a Lagrange multiplier.
The third is that it admits a straightforward and simple generalization to $\textsf{SU}(N)$.  That generalization involves adding additional
boson oscillators -- $N$ in all for $\textsf{SU}(N)$ -- which we write as $b\nd_\mu$ with $\mu=1,\ldots,N$.  The generators of $\textsf{SU}(N)$ are then
\be
S\nd_{\mu\nu}=b\yd_\mu b\nd_\nu\ .
\ee
These satisfy the $\textsf{SU}(N)$ commutation relations
\be
\big[ S\nd_{\mu\nu}\,,\,S\nd_{\mu'\nu'}\big]=S\nd_{\mu\nu'}\,\delta\nd_{\mu'\nu} - S\nd_{\mu'\nu}\,\delta\nd_{\mu\nu'}\ .
\ee
The constraint is then
\be
\sum_{\mu=1}^N b\yd_\mu b\nd_\mu= n\nd_b\ ,
\label{const}
\ee
which specifies the representation of $\textsf{SU}(N)$.  The corresponding Young tableau is one with $n\nd_b$ boxes in a single row.

\subsection{Bipartite Antiferromagnet} 
We consider the case of nearest neighbor \textsf{SU}(2) antiferromagnet, with  
interaction strength $J>0$, on a bipartite lattice with sublattices $A$ and $B$.
A bond $\ij$ is defined such that $i\in A$ and $j\in B$. 
The antiferromagnetic bond operator is defined as 
\be
{\cA}\nd_{ij}=a\nd_i b\nd_j -b\nd_i a\nd_j .
\label{15.8}
\ee
This is antisymmetric under interchange of the site indices $i$ and $j$, and transforms as a singlet
under a global \textsf{SU}(2) rotation.

Consider now a rotation by $\pi$ about the $y$ axis on sublattice $B$ only, which sends
\be
a\nd_j\to -b\nd_j\qquad,\qquad b\nd_j\to a\nd_j\ .
\label{15.9}
\ee
This  is a canonical transformation which preserves the constraint (\ref{const}). The
antiferromagnetic bond operator takes the form
\be
{\cA}\nd_{ij}\longrightarrow a\nd_i a\nd_j +b\nd_i b\nd_j.
\label{15.10}\ee
The \textsf{SU}(2) Heisenberg model is written in the form
\bea
\cH &=&J\sum_{\ij} \bS_i \cdot \bS_j\nonumber\\
&=&-{J\over 2} \sum_{\ij}\left(\cA\yd_{ij} \cA\nd_{ij}-2S^2\right).
\label{15.7}
\eea

The extension to $\textsf{SU}(N)$ for $N>2$ is straightforward.  With $N$ species of
bosons, (\ref{15.10}) generalizes to
\be
\cA\nd_{ij}~= \sum_{\mu=1}^N b\nd_{i\mu} b\nd_{j\mu}.
\label{15.11}
\ee
The nearest-neighbor  $\textsf{SU}(N)$ antiferromagnetic Heisenberg model is then
\bea
\cH&=& -{J\over N} \sum_{\ij}  \left(\cA\yd_{ij} \cA\nd_{ij}-NS^2\right)\nonumber\\
~&=&{J\over N}
\sum_{\ij} \left(\sum_{\mu,\nu} S^{\mu\nu}_i {\widetilde S}^{\nu\mu}_j -NS^2\right),
\label{15.12}
\eea
where 
\be
{\widetilde S}^{\mu\nu}_j = -\,b\yd_{j\nu} b\nd_{j\mu}
\ee
are the generators of the {\em conjugate
representation} \index{conjugate representation}on sublattice $B$.
One should note that $\cH$ of (\ref{15.12}) 
is not invariant under uniform $\textsf{SU}(N)$
transformations $U$ but only under staggered conjugate rotations $U$ and $U^\dagger$  on
sublattices $A$ and $B$, respectively.

\subsection{Non-bipartite (Frustrated) Antiferromagnets} 
For the group \textsf{SU}(2), one can always form a singlet from two sites in an identical spin-$S$ representation.
That is, the tensor product of two spin-$S$ states always contains a singlet:
\be
S\otimes S=0\oplus 1\oplus\cdots\oplus 2S \ .
\ee
For $\textsf{SU}(N)$ this is no longer the case.  For example, for two \textsf{SU}(3) sites in the fundamental
representation, one has ${\bf 3}\otimes {\bf 3}={\overline{\bf 3}}\oplus {\bf 6}$.  One needs three
constituents to make an \textsf{SU}(3) singlet, as with color singlets in QCD, and $N$ constituents
in the case of $\textsf{SU}(N)$.   This is why, in the case of the antiferromagnet, one chooses the conjugate
representation on the B sublattice -- the product of a representation and its conjugate always contains
a singlet.

But what does one do if the lattice is not bipartite?  This situation was addressed by Read and Sachdev
\cite{SPN}, who extended the Schwinger boson theory to the group $\textsf{Sp}(N)$.  This amounts to generalizing
the link operator $\cA_{ij}$ in (\ref{15.8}) to include a flavor index $m$:
\bea
\cA_{ij}&=& \sum_{m=1}^N  \left(a_{im}b_{jm} - b_{im}a_{jm} \right) \nonumber\\
&\equiv&\sum_{\alpha,\alpha'=1}^{2N} \Lambda\nd_{\alpha\alpha'}\,b\nd_{i\alpha} \,b\nd_{j\alpha'}\ .
\eea
Here, the indices $\alpha$ and $\alpha'$ run from $1$ to $2N$.  They may be written
in composite form as $\alpha\to (m,\mu)$, where $m$ runs from $1$ to $N$ and
$\mu$ from $1$ to $2$ (or $\uar$ and $\dar$).  In this case, on each site one
has $b\nd_{m\uar}=a\nd_m$ and $b\nd_{m\dar}=b\nd_m$.  The matrix
$\Lambda\nd_{\alpha\alpha'}$ is then $\Lambda\nd_{m\mu,n\nu}=\delta\nd_{mn}\,
\epsilon\nd_{\mu\nu}$, where $\epsilon\nd_{\mu\nu}=i\sigma^y_{\mu\nu}$
is the rank two antisymmetric tensor.

If we make a global transformation on the Schwinger
bosons, with $b_{i\alpha}\to U_{\alpha\alpha'}\,b_{i\alpha'}$, then we find
\be
\cA\nd_{ij}\to (U^{\rm t}\! \Lambda U)\nd_{\alpha\alpha'}\,b\nd_{i\alpha}\,b\nd_{j\alpha'}\ .
\ee
Thus, the link operators remain invariant under the class of complex transformations which satisfy $U^{\rm t}\! \Lambda U=\Lambda$.
This is the definition of the group ${\textsf{Sp}}(2N,{\mathbb C}$).   If we further demand that $U\in{\textsf U}(2N)$, which is necessary
if the group operations are to commute with the total occupancy constraint, we arrive at the group
\be
{\textsf{Sp}}(N)={\textsf{Sp}}(2N,{\mathbb C})\cap {\textsf U}(2N)\ .
\ee
For $N=1$ one has $\textsf{Sp}(1)\simeq\textsf{SU}(2)$.
The particular representation is again specified by the local boson occupation,
$n\nd_b=\sum_\alpha b\yd_{i\alpha} b\nd_{i\alpha}$.
The Hamiltonian is
\be
\cH= -{1\over 2N} \sum_{i<j} J\nd_{ij} \,\cA\yd_{ij} \cA\nd_{ij}\ .
\ee
Here, we have allowed for further neighbor couplings, which can be used to introduce frustration in the square lattice antiferromagnet,
{\it e.g.\/} the $J_1-J_2-J_3$ model \cite{SPN}.  For each distinct coupling $J\nd_{ij}$ (assumed translationally invariant), a new 
Hubbard-Stratonovich decomposition is required.

One can also retain the definition in (\ref{15.11}) even for frustrated lattices.  In this case, under a global
transformation $b\nd_{i\mu}\to U\nd_{\mu\nu} \,b\nd_{i\nu}$, the link operator $\cA\nd_{ij}$ transforms as
$\cA\nd_{ij}\to (U^{\rm t} U)\nd_{\mu\nu}\, b\nd_{i\mu} b\nd_{j\nu}$, and invariance of $\cA\nd_{ij}$ requires $U^{\rm t} U=1$.
This symmetry is that of the complex orthogonal group $\textsf{O}(N,{\mathbb C})$.  Once again, we require $U\in \textsf{SU}(N)$
so that the constraint equation remains invariant.  We then arrive at the real orthogonal group
$\textsf{O}(N)=\textsf{O}(N,{\mathbb C})\cap\textsf{SU}(N)$.  For $N=2$, in terms of the original spin operators, we have
\bea
-{1\over N} \cA\yd_{ij}\cA\nd_{ij}&=&-{1\over N}\,b\yd_{i\mu} b\nd_{i\nu} b\yd_{j\mu} b\nd_{j\nu}\\
&=&-\big(S^x_i S^x_j - S^y_i S^y_j + S^z_j S^z_j +S^2)\qquad \big(N=2\big)\ .
\eea
On a bipartite lattice, one can rotate by $\pi$ about the $y$-axis on the $B$ sublattice, which recovers the
isotropic Heisenberg interaction ${\bf S}_i\cdot{\bf S}_j$.  On non-bipartite lattices, the $N=2$ case does not
correspond to any isotropic \textsf{SU}(2) model, and so one loses contact with the original problem. 

\section{Mean Field Hamiltonian}
Within a functional integral approach, one introduces a single real field $\lambda\nd_i(\tau)$ on each site to
enforce the occupancy constraint, and a complex Hubbard-Stratonovich field $Q\nd_{ij}(\tau)$ on each link
to decouple the interaction.  At the mean field level it is assumed that these fields are static.  This results
in the mean field Hamiltonian
\bea
\cH^\ssr{MF}&=&{pN\over J}\sum_{i<j} |Q\nd_{ij}|^2 + \sum_{i<j} \big(Q\nd_{ij}\,\cA\yd_{ij} + Q^*_{ij}\,\cA\nd_{ij}\big)\\
&&\qquad + \sum_i\lambda\nd_i\big(b\yd_{i\alpha}b\nd_{i\alpha} -n\nd_b\big) +(\cV N)^{-1/2}
\sum_{i\alpha} \big(\phi^*_{i\alpha} b\nd_{i\alpha} + \phi\nd_{i\alpha} b\yd_{i\alpha}\big)\ ,\nonumber
\eea
where $\cV$, the volume, is the number of Bravais lattice sites, and where $\alpha$ runs from $1$ to
$N$ for the $\textsf{SU}(N)$  models (for which $p=1$), and from $1$ to $2N$ for the $\textsf{Sp}(N)$ models (for which $p=2$).
The field $\phi\nd_{i\alpha}$, which couples linearly to the Schwinger bosons, is conjugate to the condensate parameter
$\langle b\yd_{i\alpha}\rangle$, which means
\be
{\partial F\over\partial \phi^*_{i\alpha}}={\langle b_{i\alpha}\rangle\over\sqrt{\cV N}}\ .
\ee

Let us further assume that the mean field solution has the symmetry of the underlying lattice, and that the interactions are only
between nearest neighbor sites on a Bravais lattice.  Then, after Fourier transforming, we have
\bea
\cH^\ssr{MF}&=&\cV  N \bigg({pz\over 2J} |Q|^2 - {\na\over N}\lambda\bigg)+{z\over 2}\sum_{\bk,\alpha,\alpha'}\Big[ Q\,\xk\,
K\nd_{\alpha\alpha'}\,b\yd_{\bk,\alpha} b\yd_{-\bk,\alpha'}
+ Q^*\, \xks\, K\nd_{\alpha\alpha'}\,b\nd_{\bk,\alpha} b\nd_{-\bk,\alpha'}\Big]\nonumber\\
&&\qquad+\lambda\sum_{\bk,\alpha} b\yd_{\bk,\alpha} b\nd_{\bk,\alpha}+(\cV N)^{-1/2}\sum_{\bk,\alpha}
\big(\phi^*_{\bk,\alpha} \,b\nd_{\bk,\alpha} + \phi\nd_{\bk,\alpha} \,b\yd_{\bk,\alpha}\big)\ ,
\eea
where $z$ is the lattice coordination number, and
where $K\nd_{\alpha\alpha'}=\delta\nd_{\alpha\alpha'}$ for $\textsf{SU}(N)$ and
$K\nd_{\alpha\alpha'}=\Lambda\nd_{\alpha\alpha'}$ for $\textsf{Sp}(N)$.  We define
\be
\xk\equiv {2\over z}\,{\sum_{\delta}}' \eta\nd_\delta\,e^{i\bk\cdot\delta}\ ,
\ee
where the sum is over all {\it distinct\/} nearest neighbor vectors in a unit cell.  That is, $-\delta$ is not included in the sum.  
The quantity $\eta\nd_\delta=\pm 1$ is a sign about which we shall have more to say presently.  On the square lattice,
for example, $\xk=\eta\nd_x \,e^{ik\nd_x} + \eta\nd_y\,e^{ik\nd_y}$.  For symmetric $K\nd_{ab}$, owing to the sum on $\bk$, we can replace
$\xk$ with its real part, while for antisymmetric $K\nd_{ab}$ we must replace it with $i$ times its imaginary part.  We therefore define
\bea
\gk&=&{2\over z}\,{\sum_\delta}'\eta\nd_\delta\,\cos(\bk\cdot\delta) \quad {\rm if}\ K=K^{\rm t} \\
&=&{2i\over z}\,{\sum_\delta}'\eta\nd_\delta\,\sin(\bk\cdot\delta) \quad {\rm if}\ K=-K^{\rm t}
\eea
The sign $\eta\nd_\delta$ is irrelevant on bipartite lattices, since it can be set to unity for all $\delta$ simply by choosing
an appropriate center for the Brillouin zone.  But on frustrated lattices, the signs matter.

It is now quite simple to integrate out the Schwinger bosons.  After we do so, we make a Legendre transformation to replace the field $\phi\nd_{i\alpha}$ with the order parameter
$\beta\nd_{i\alpha}=\langle b\nd_{i\alpha}\rangle/\sqrt{\cV N}$,
by writing
\be
G=F-\sum_{i\alpha}\big(\phi\nd_{i\alpha}\,\beta^*_{i\alpha}+ \phi^*_{i\alpha}\,\beta\nd_{i\alpha}\big)\ .
\ee
The final form of the free energy per site, per flavor, is
\be
g\equiv {G\over \cV  N}={pz\over 2J} |Q|^2 -\big(\kappa+\frac{p}{2}\big) \lambda + p\!\int\limits_\ssr{BZ}\!\!{d^d\!k\over (2\pi)^d}
\bigg[\frac{1}{2}\ok + T\,\ln\Big( 1 - e^{-\ok/T}\Big)\bigg] + E\nd_{\rm con}\ ,
\ee
where $\kappa=\na/N$, and $E\nd_{\rm con}$ is the condensation energy,
\be
E\nd_{\rm con}=\lambda \sum_{\bk,\alpha} |\beta\nd_{\bk,\alpha}|^2 + \frac{1}{2}z\!\sum_{\bk,\alpha,\alpha'} K\nd_{\alpha\alpha'} 
\Big( Q\,\gk\,\beta^*_{\bk,\alpha}\beta^*_{-\bk,\alpha'} +  Q^*\,\gks\,\beta\nd_{\bk,\alpha}\beta\nd_{-\bk,\alpha'}\Big)\ .
\ee
The dispersion is given by
\be
\ok=\sqrt{\lambda^2-\big|z Q \gk\big|^2}\ .
\ee
The fact that $g$ is formally of order $N^0$ (assuming $\kappa$ is as well) allows one to generate a systematic
expansion of the free energy in powers of $1/N$. 

\subsection{Mean Field Equations}
The mean field equations are obtained by extremizing the free energy $G$ with respect to the parameters $\lambda$, $Q$, and 
$\beta\nd_{\bk,a}$.  Thus,
\bea
\kappa+\frac{p}{2}&=& p\!\int\limits_\ssr{BZ}\!\!{d^d\!k\over (2\pi)^d}\>{\lambda\over\omega_\bk}\,
\big(n\nd_\bk(T)+\frac{1}{2}\big) + \sum_{\bk,\alpha}  |\beta\nd_{\bk,\alpha}|^2 \label{MFA}\\
{pz\over J}\,|Q|^2&=&p\!\int\limits_\ssr{BZ}\!\!{d^d\!k\over (2\pi)^d}\>{\big|z Q\gk|^2\over\omega_\bk}\,\big(n\nd_\bk(T)+\frac{1}{2}\big)
+\lambda \sum_{\bk,\alpha}  |\beta\nd_{\bk,\alpha}|^2 \label{MFB}\\
0&=&\lambda \beta\nd_{\bk,\alpha} +z Q \gk \sum_{\alpha'} K\nd_{\alpha\alpha'}\,\beta^*_{-\bk,\alpha'}\ .\label{MFC}
\eea
Here, $n\nd_\bk(T)=\big(e^{\ok/T}-1\big)^{-1}$ is the thermal Bose occupancy function.
In deriving the second of the above mean field equations, we have also invoked the third.
Assuming that the condensate occurs at a single wavevector $\bk$, the last equation requires that $\ok=0$ at the ordering wavevector,
ensuring gaplessness of the excitation spectrum.  When there is no condensate, $\beta\nd_{\bk,\alpha}=0$ for all $\bk$ and $\alpha$.

It is instructive to compute
$\langle\, S^{\alpha\alpha'}_\bR\,\rangle=\big\langle\, b\yd_{\bR,\alpha}\,b\nd_{\bR,\alpha'} -
\kappa\,\delta\nd_{\alpha\alpha'}\,\big\rangle$, which serves as the local order parameter.
After invoking the mean field equations, one finds
\be
\big\langle\, S^{\alpha\alpha'}_\bR\,\big\rangle = N\,\sum_{\bk,\bk'} e^{i(\bk'-\bk)\cdot\bR}\,
\beta^*_{\bk \alpha}\,\beta\nd_{\bk'\alpha'} -\sum_{\bk,\alpha''} \big|\beta\nd_{\bk \alpha''}\big|^2\,
\delta\nd_{\alpha\alpha'}\ .
\ee
Note that the trace of the above expression vanishes on average ({\it i.e.\/} upon summing
over $\bR$), and vanishes locally provided that the condensate satisfies the orthogonality
condition
\be
\sum_\alpha\beta^*_{\bk\alpha}\,\beta\nd_{\bk'\alpha}=\delta\nd_{\bk\bk'}\sum_\alpha 
\big|\beta\nd_{\bk \alpha}\big|^2\ .
\ee
In the case of an $\textsf{SU}(N)$ antiferromagnet on a (bipartite) hypercubic lattice,
the condensate occurs only at the zone center $\bk=0$ and the zone corner $\bk=\vpi$.
One then has
\be
\big\langle\, S^{\alpha\alpha'}_\bR\,\big\rangle = N \big(\beta^*_{0\alpha}\,\beta\nd_{\vpi \alpha'}
+ \beta^*_{\vpi \alpha}\,\beta\nd_{0 \alpha'}\big)\,e^{i\vpi\cdot\bR}\ .
\ee
Thus, Bose condensation of the Schwinger bosons is equivalent to long-ranged magnetic order.

At $T=0$, there is a critical value of $\kappa$ above which condensation occurs.  To find this value, we invoke all three equations, but
set the condensate fraction to zero.   For the $\textsf{SU}(N)$ models, the minimum of the dispersion occurs at the zone center, $\bk=0$.
Setting $\omega\nd_{\bk=0}=0$, we obtain the relation $\lambda=z |Q|$.  The first equation then yields
\be
\kappa\nd_{\rm c}=\frac{1}{2}\! \int\limits_\ssr{BZ}\!\!{d^d\!k\over (2\pi)^d}\>\big(1-|\gk|^2\big)^{-1/2}-\frac{1}{2}\ .
\ee
For $d=1$, there is no solution, and there is never a condensate.  For $d=2$, one finds $\kappa\nd_{\rm c}=0.19$ on the square lattice \cite{AA}.
Since $\kappa=S$ for the \textsf{SU}(2) case, this suggests that even the minimal $S=\frac{1}{2}$ model is N{\'e}el ordered on the square lattice,
a result which is in agreement with quantum Monte Carlo studies. 

Consider next the $\textsf{Sp}(N)$ model on the triangular lattice.  We first must adopt a set of signs $\eta\nd_\delta$.  There are three bonds $\delta\nd_{1,2,3}$
per unit cell, along the directions ${\bf a}\nd_1$, ${\bf a}\nd_2$, and ${\bf a}\nd_1-{\bf a}\nd_2$, where the primitive
direct lattice vectors are ${\bf a}\nd_1=a\,{\hat{\bf x}}$ and ${\bf a}\nd_2=\frac{1}{2} a\,{\hat{\bf x}}+\frac{\sqrt{3}}{2} a\,{\hat{\bf y}}$.
Lattice symmetry suggests $\eta\nd_1=+1$, $\eta\nd_2=-1$, and $\eta\nd_3=+1$ (as opposed to all $\eta\nd_\delta=1$), resulting in \cite{KAG}
\be
\gk=\frac{1}{3}\sin \theta\nd_1 - \frac{1}{3}\sin \theta\nd_2 + \frac{1}{3}\sin (\theta\nd_2-\theta\nd_1) \ ,
\ee
where the wavevector is written as
\be
\bk = {\theta\nd_1\over 2\pi}\,{\bf G}\nd_1 + {\theta\nd_2\over 2\pi}\,{\bf G}\nd_2\ ,
\ee
with ${\bf G}\nd_{1,2}$ being the two primitive reciprocal lattice vectors for the triangular lattice.  The maximum of $|\gk|^2$,
corresponding to the minimum of the dispersion $\ok$, occurs when $\gk$ lies at one of the two inequivalent zone corners.
In terms of the $\theta\nd_i$, these points lie at $(\theta\nd_1,\theta\nd_2)=(\frac{4\pi}{3},\frac{2\pi}{3})$, where $\gk=-\frac{\sqrt{3}}{2}$,
and at $(\theta\nd_1,\theta\nd_2)=(\frac{2\pi}{3},\frac{4\pi}{3})$, where $\gk=\frac{\sqrt{3}}{2}$.  Sachdev \cite{KAG} has found $\kappa\nd_{\rm c}=0.34$
for the triangular structure.  As one would guess, frustration increases the value of $\kappa\nd_{\rm c}$ relative to that on the square lattice.
On the Kagom{\'e} lattice, which is even more highly frustrated, he finds $\kappa\nd_{\rm c}=0.53$.

\section{The Mean Field Antiferromagnetic Ground State} 
For a finite system (no long range order or Bose condensation) one can explicitly write down the
 ground state of the $\textsf{SU}(N)$ Schwinger Boson Mean Field Theory  $ \Psi^{MF} $. It   is simply the vacuum of all the Bogoliubov operators 
 \be
\beta\nd_{\bk,\alpha} {\rm\Psi}^\ssr{MF} =0\qquad\forall \ \bk,\alpha.
\ee
where,  
 \be
\beta\nd_{\bk \alpha }= \cosh\theta\nd_\bk\,  b\nd_{\bk \alpha } -
\sinh\theta\nd_\bk\,  b\yd_{-\bk \alpha},
\label{16.31}
\ee
and 
\be
\tanh 2\theta_\bk = - {zQ\gk\over \lambda}.\nonumber\\
\label{16.34}
\ee

The ground state wavefunction  ${\rm\Psi}^\ssr{MF}$ can be explicitly
written in terms of the original Schwinger bosons as
\bea
{\rm\Psi}^\ssr{MF}\!&=& C  \exp
\left[\half \sum_{ij}u\nd_{ij} \sum_m b^\dagger_{i a} b^\dagger_{j a} 
\right]|0\rangle~,
\nonumber\\
u\nd_{ij}&=&{1\over \cV } \sum_\bk e^{ i\bk\bR\nd_{ij}}\tanh\theta\nd_\bk.
\label{16.41.3}
\eea
For   $N=2$, using the {\em unrotated} operators $a^\dagger$ and $b^\dagger$,
the mean field Schwinger boson ground state ${\rm\Psi}^\ssr{MF}$ is
\be
{\rm\Psi}^\ssr{MF}_{N=2}= \exp
\left[ \sum_{i\in A \atop j\in B} u\nd_{ij}\left( a^{\dag}_{i}b^{\dag}_{j}-b^{\dagger}_{i}a^{\dagger}_{j}\right)\right]|0\rangle
.\label{16.36}
\ee 
${\rm\Psi}^\ssr{MF}$ contains many configurations with occupations different from $2S$
and is therefore not a pure spin state.
As shown in \cite{Raykin}, under Gutzwiller projection \index{Gutzwiller projection} it
reduces to a valence bond state.
Since $\tanh\theta\nd_{\bk+\vpi}=-\tanh\theta\nd_{\bk}$,
where  $\vpi=(\pi,\pi,\ldots)$, the bond parameters $u_{ij}$ only connect sublattice $A$ to $B$.   Furthermore, one can verify that for the nearest neighbor model above, 
$u\nd_{ij} \ge 0$, and therefore
the  valence bond states obey  Marshall's sign. 

Although ${\rm\Psi}^\ssr{MF}$ is manifestly rotationally invariant, it may or may not 
exhibit long-ranged antiferromagnetic (N{\'e}el) order.  This depends on the long-distance 
decay of $u\nd_{ij}$. As was shown \cite{Raykin,Havilio} the SBMFT ground state for the nearest neighbor
model is disordered in one dimension, and can exhibit long-range order in two dimensions
for physically relevant values of $S$.

\begin{figure}[t]
\centering
\includegraphics*[width=.7\textwidth]{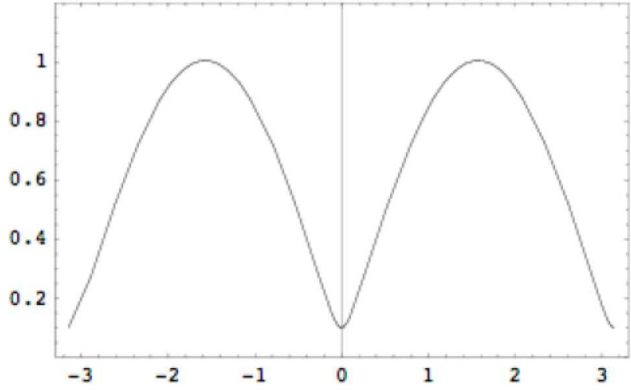}
\caption{
\label{oned} Mean field dispersion $\omega_{k}$, 
in the domain $-\pi<k<\pi$, for the one-dimensional
antiferromagnet.} 
\end{figure}

For further calculations, it is convenient to introduce the   parametrizations:
\bea
\ok  &\equiv & c\, \sqrt{(2\xi)^{-2}  + {z\over 2}(1-\gamma_\bk^2)}\nonumber\\
c  &\equiv &   \sqrt{2z}\>  |Q| \nonumber\\
\xi^{-1} &\equiv &{2\over c} \sqrt{\lambda^2 - \big(z|Q|\big)^2}\nonumber\\
t~&=&{T\over z|Q|} \ .
\label{16.36.1}
\eea
Here, $c$, $\xi$, and $t$  describe the spin wave velocity, correlation length,
and the dimensionless temperature, respectively.  In Fig. \ref{oned} the
dispersion for the one-dimensional antiferromagnet is drawn. 

At the zone center and zone corner the
mean field dispersion is that of free massive relativistic bosons,
\be
\ok \approx c\,\sqrt{(2\xi )^{-2}+ |\bk-\bk_\gamma |^2}~,~~~~~\bk_\gamma= 0,\vpi.
\ee
When the  gap (or ``mass'' $c/2\xi $) vanishes, $\ok$ are Goldstone modes which reduce  to dispersions of  antiferromagnetic spin waves.

\section{Staggered Magnetization in the Layered Antiferromagnet}
Consider now a layered antiferromagnet on a cubic lattice where the in-plane nearest
neighbor coupling is $J$ and the interlayer coupling is $\alpha J$, with $\alpha\ll 1$.
We expect  long range magnetic order at a finite N\'eel temperature $T_{\rm N}$. The
order parameter, which is the staggered magnetization ${\bf M}=\langle (-1)^l\,
e^{i\vpi\cdot\bR}\,{\bf S}\nd_{\bR,\,l}\rangle$ becomes finite
when the in-plane correlation length $\xi$, which diverges exponentially at low $T$, produces an effective coupling between neighboring layers $\alpha\, \xi^2(T\nd_{\rm N})$
which is of order unity.  Here $\bR$ locates the site within a plane, and $l$ is the
layer index.  This means in effect that the coarse grained spins start to
interact as if in an isotropic three dimensional cubic lattice which orders at
$T_{\rm N}$. 
The interlayer mean field theory, introduced by Scalapino, Imry and Pincus (SIP) \cite{SIP} in the 1970's, can be applied within the SBMFT.
Here we follow Keimer {\em et al.}\cite{Keimer}, and Ofer {\em et al.}\cite{Ofer}, to compute the temperature dependent staggered magnetization,  
in the range $T\in [0,T\nd_{\rm N}]$. 

The Hamiltonian is given by
\be
\cH= \sum_{\bR,l} \left( \bS\nd_{\bR,\,l} \cdot \bS\nd_{\bR+\bxh,\,l}+
 \bS\nd_{\bR,\,l} \cdot \bS\nd_{\bR+\byh,\,l} + \alpha\, 
 \bS\nd_{\bR,\,l} \cdot \bS\nd_{\bR,\,l+1}\right)
\ee
The interplane coupling is decomposed using  Hartree-Fock  staggered magnetization field:
\be
 \alpha \, S^{z}_{\bR,\,l} \,S^z_{\bR,\,l+1}  \longrightarrow (-1)^l\,e^{i\vpi\cdot\bR}
  \left( S^{z}_{\bR,\,l}  - S^z_{\bR,\,l+1}\right)\,h\nd_{\bR,l} 
  - {h_{\bR,l}^2 \over \alpha }\ ,\label{HFS}
\ee
where it is assumed that ${\bf M}=M\,\bzh$.  Here $h\nd_{\bR,l}$ is the local
N{\'e}el field due to any ordering in the neighboring layers.  Assuming a uniform
solution, $h\nd_{\bR,l}=h$, self-consistency is achieved when
\be
{h\over \alpha}= 2 \, M(T,h)=\langle\,a\yd_\bR\,a\nd_\bR\,\rangle
-\langle\,b\yd_\bR\,b\nd_\bR\,\rangle\ ,
\label{SelfCons}
\ee
where $M(T,h)$ is the staggered magnetization response to an ordering staggered field
$h$ on a single layer.

Extracting $T\nd_{\rm N}$ is relatively easy, since  as $T\to T\nd_{\rm N}$,
$h\to 0$, and the expressions for $\langle\,a\yd_\bR\,a\nd_\bR\,\rangle$
and $\langle\,b\yd_\bR\,b\nd_\bR\,\rangle$.  In this limit, one finds that the
second mean field equation in (\ref{MFB}) is not affected by the staggered field,
which simplifies the calculation.  At $T\nd_{\rm N}$ one finds
\be
2\alpha\, \chi^{\rm s}_\ssr{2D}(T\nd_{\rm N}) =1\ .
\ee
Since we know that $\chi^{\rm s}_\ssr{2D} \propto  \xi_\ssr{2D}^2(T\nd_{\rm N})$,
we recover  the ordering temperature of the SIP theory.
The more precise calculation yields (restoring the Heisenberg exchange energy scale $J$),
\be
T\nd_{\rm N}={ J \over \log\alpha}~\left({2\pi M_0 \over  |\log\left(\frac{1}{4}\pi^2
\log (4\alpha/\pi)\,M_0^2\right)| }\right)
\ee
The numerically determined  $M(T)=h(T)/2\alpha$  is shown in Fig.\ref{smag}
for various anisotropy parameters.
\begin{figure}[t]
\centering
\includegraphics*[width=10cm]{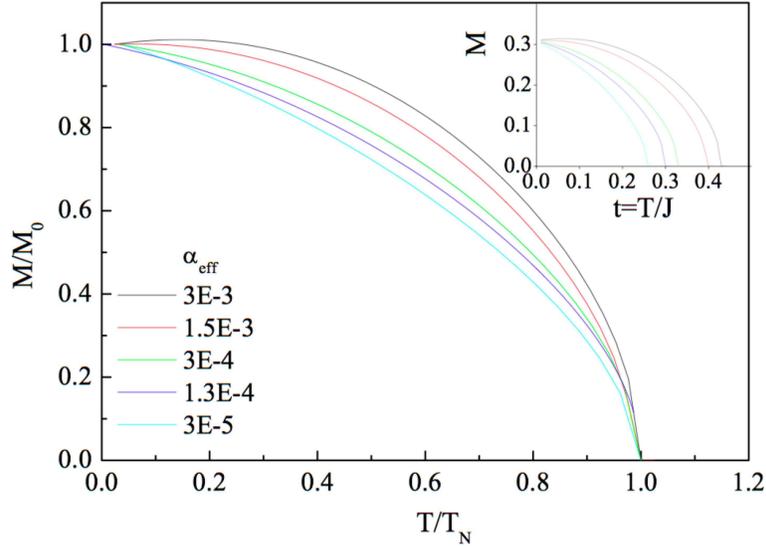}
\caption
{\label{smag} Numerical solution, from Ofer {\em et al.} \cite{Ofer},  of the staggered magnetization $M(T)$
of the layered antiferomagnet for various values of  anisotropy parameter $\alpha_{\rm eff}$. $M_0=S-0.19660$ is the 
zero temperature staggered magnetization, and $T_N$ is the N\'eel temperature.} 
\label{smag}
\end{figure}

One can also analyze the layered antiferromagnet using the SBMFT's native
decoupling scheme, without proceeding via the interlayer mean field theory
of (\ref{HFS}).  Starting from an anisotropic
Heisenberg model with in-plane exchange $\Jpa$ and interlayer exchange $\Jpe$, one 
assumes a mean field solution where $Q\nd_{ij}=\Qpa$ when $\langle ij\rangle$
is an in-plane bond, and $Q\nd_{ij}=\Qpe$ when $\langle ij\rangle$
is an out-of-plane bond.  The second mean field equation, (\ref{MFB}), then becomes
two equations.  The N{\'e}el temperature can be written $T\nd_{\rm N}=\Jpa\,
f(\Jpe/\Jpa)$, where $f(\alpha)$ is a dimensionless function.  To find $f(\alpha)$,
we demand that the spectrum be gapless, but the condensate vanishes.
This results in two coupled equations,
\bea
\kappa+\half&=&\intl\!d\gpa\,\rhpa(\gpa)\!\intl\!d\gpe\,\rhpe(\gpe)\>
{1+\epsilon\over{\rm\Omega}(\gpa,\gpe)}\,\Big(n(\gpa,\gpe)+\frac{1}{2}\Big)
\label{IMFA}\\
{\Jpe\over 2\Jpa}&=&{\intl\!d\gpa\,\rhpa(\gpa)\intl\!d\gpe\,\rhpe(\gpe)\>
(\gamma_\parallel^2/{\rm\Omega})\,(n+{1\over 2})\over
\intl\!d\gpa\,\rhpa(\gpa)\intl\!d\gpe\,\rhpe(\gpe)\>
(\gamma_\perp^2/{\rm\Omega})\,(n+{1\over 2})}\ ,\label{IMFB}
\eea
where $\epsilon=\big|\Qpe/2\Qpa\big|$ and ${\rm\Omega}(\gpa,\gpe)=
\sqrt{(1+\epsilon)^2-(\gamma\nd_\parallel+\epsilon\,\gamma\nd_\perp)^2}\>$, and where
$n=\big(e^{{\rm\Omega}/t_{\rm c}}-1\big)^{-1}$, with $t\nd_{\rm c}=
T\nd_{\rm N}/4\, |\Qpa|$.  Once the above two equations are solved for $\epsilon$
and $t\nd_{\rm c}$, we determine $\Qpa$ from
\be
\Qpa=\Jpa\intl\!d\gpa\,\rhpa(\gpa)\!\intl\!d\gpe\,\rhpe(\gpe)\>
{\gamma_\parallel^2\over{\rm\Omega}}\,\Big(n+{1\over 2}\Big)\ .
\ee
The functions $\rhpa$ and $\rhpe$ are given by
\be
\rhpa(\gpa)={2\over\pi^2}\,{\rm K}\big(1-\gamma_\parallel^2\big)\qquad,\qquad
\rhpe(\gpe)={1\over \pi\sqrt{1-\gamma_\perp^2}}\ .
\ee

{\em Acknowledgements.}
We are grateful to A. Keren for help in preparing these notes. 
Support from the Israel Science Foundation, the US Israel Binational Science Foundation,  and the fund for Promotion of Research at Technion is acknowledged.
We are grateful for the hospitality of Aspen Center for Physics, and Lewiner Institute for Theoretical Physics, Technion,  where these notes were written.

\printindex
\end{document}